\documentclass{article}
\usepackage{spconf,amsmath,graphicx}
\usepackage{hyperref, booktabs, multirow}
\usepackage[subtle]{savetrees}

\title{Improving the Speaker Anonymization Evaluation's Robustness to Target Speakers with Adversarial Learning}
\name{
Carlos Franzreb\textsuperscript{1}, Arnab Das\textsuperscript{1}, Tim Polzehl\textsuperscript{1}, Sebastian Möller\textsuperscript{2}
\thanks{Funded by Federal Ministry of Education and Research, Germany (BMBF 16KIS2048).}
}
\address{
\textsuperscript{1} DFKI, Germany,
\textsuperscript{2} Technical University of Berlin, Germany
\\ \url{carlos.franzreb@dfki.de}
}
\begin{document}
%\ninept
%
\maketitle
\begin{abstract} % 100 to 150 words
The current privacy evaluation for speaker anonymization often overestimates privacy when a same-gender target selection algorithm (TSA) is used, although this TSA leaks the speaker's gender and should hence be more vulnerable.
We hypothesize that this occurs because the evaluation does not account for the fact that anonymized speech contains information from both the source and target speakers.
To address this, we propose to add a target classifier that measures the influence of target speaker information in the evaluation, which can also be removed with adversarial learning.
Experiments demonstrate that this approach is effective for multiple anonymizers, particularly when using a same-gender TSA, leading to a more reliable assessment.
\end{abstract}
\section{Introduction}
\label{sec:intro}

% spkanon
Given an utterance, speaker anonymization aims to conceal the source speaker's identity while keeping the linguistic and paralinguistic content intact \cite{jin_speaker_2009}.
This is achieved by replacing the source speaker with a target speaker (\textit{source} and \textit{target} for short), following the approach for voice conversion \cite{sisman_overview_2021}.
% Nature of anonymized speech
If the anonymization is imperfect, as is usually the case, the source speaker will still be present in the anonymized speech, which will thus comprise information of both the source and the target.
This dichotomy is not addressed in the current privacy evaluation.
We hypothesize that this is the reason why the evaluation sometimes overestimates privacy: it is sensitive to targets because it is not informed about them.

% Privacy evaluation
The privacy evaluation was designed by the VoicePrivacy initiative \cite{tomashenko_introducing_2020}.
It trains a speaker recognizer with anonymized speech.
% Target selection
How targets are selected to anonymize the utterances can influence the privacy estimation, although different targets provide in principle the same amount of privacy.
The evaluation behaves erratically when the target selection algorithm (TSA) used to anonymize the training data does not ensure unlinkability between sources and targets \cite{champion_anonymizing_2023}.
Also, if the TSA always assigns the same target to all the utterances of each source, the recognizer will learn the assignments instead of trying to identify the sources \cite{panariello_risks_2025}.
The recognizer's learning strategy works for the training data, where all speakers are seen, but will afterward fail to identify the unseen anonymized speakers during the evaluation, where the target assignments are unknown.
All these results suggest the recognizer's sensitivity to the TSA is a bug of the evaluation, not a feature of the anonymizer.

% Same-gender selection
The privacy of the anonymizer \textit{private kNN-VC} \cite{franzreb_private_2025} is overestimated when a same-gender constraint is placed on the TSA, although the utterances of each source are anonymized with different targets and unlinkability is ensured by randomly picking a target from those that share the source's gender.
The evaluation's outcome suggests that private kNN-VC provides perfect privacy, whereas the outcome is worse when the same-gender constraint is removed and targets are picked randomly.
In theory, we expect the opposite: the same-gender constraint means that the recognizer can easily infer the source's gender, which should boost its ability to recognize speakers.
% Use cases
Preserving gender throughout anonymization is a requirement for many use cases (e.g. medical consultations \cite{franzreb_towards_2024}), as it provides useful context for understanding the anonymized speech.
Given that it poses a bigger challenge from the perspective of the anonymizer, as the gender is not anonymized, our goal is to evaluate it properly.

% Contribution
% Not accounting for targets in the evaluation
In the current privacy evaluation \cite{tomashenko_introducing_2020}, the recognizer is not informed that the speech is anonymized, i.e. that it inherits aspects from several speakers (sources and targets).
The recognizer is trained with anonymized speech labeled with the sources, but it has no information about which target was used to anonymize each utterance. 
We expect the recognizer to discard target information and focus on identifying sources.
As the aforementioned related work shows \cite{champion_anonymizing_2023, panariello_risks_2025, franzreb_private_2025}, the evaluation fails to meet these expectations for some TSAs, leading to an overestimation of privacy.

% Scope
Our study\footnote{All the necessary code and information to reproduce our experiments is available on GitHub: \url{https://github.com/carlosfranzreb/spane}.} investigates the privacy overestimation that occurs when a same-gender TSA is used by adding a target classifier to the recognizer, besides the source classifier which it already has during training.
The target classifier tells us how much target information the recognizer is encoding, and also allows us to remove this information from the recognizer's output with adversarial learning.
According to our results, doing so improves the evaluation's performance for two distinct anonymizers when the same-gender TSA is used.

\begin{figure}[th]
    \centering
    \includegraphics[width=1.1\linewidth]{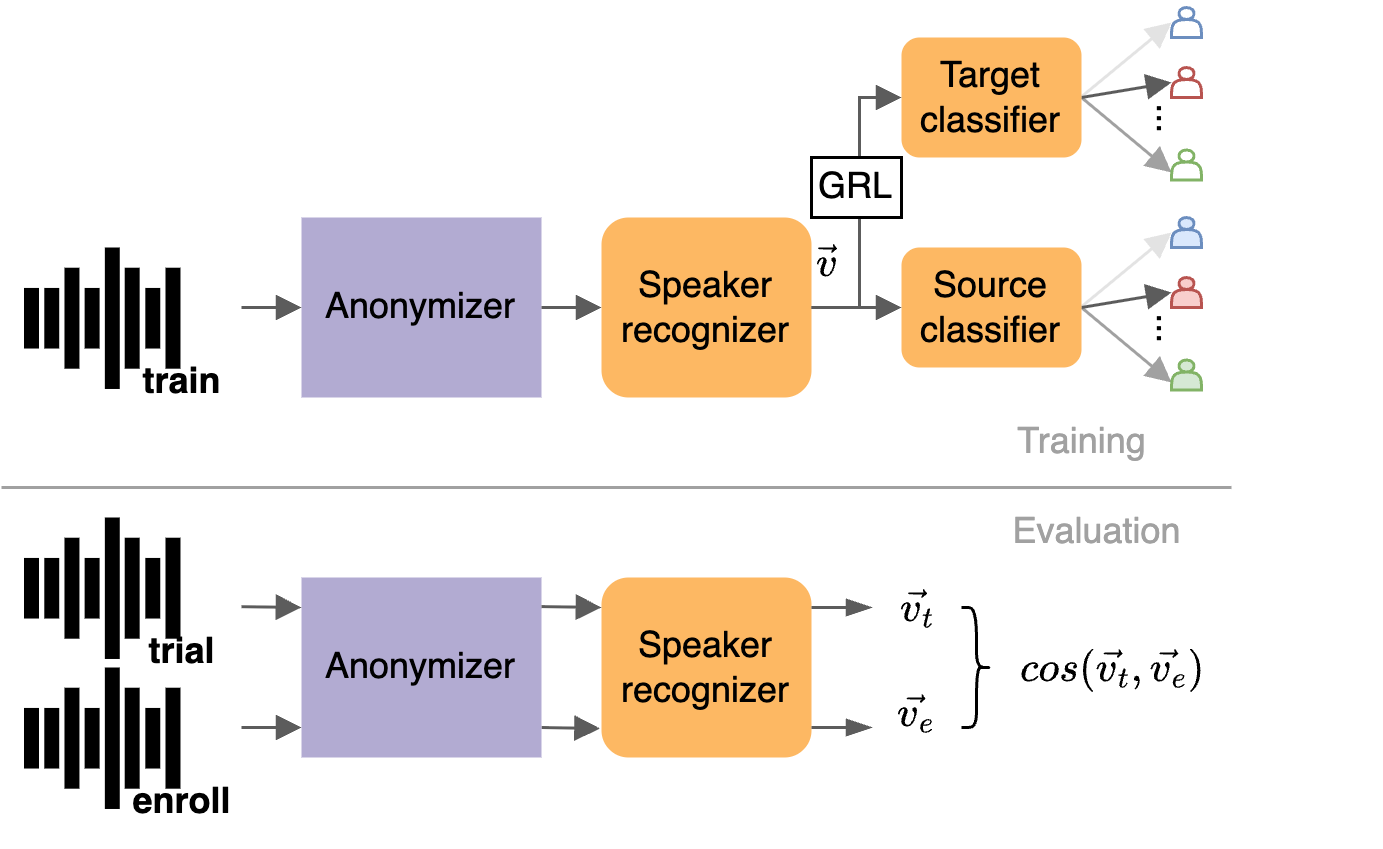}
    \caption{Diagram of the privacy evaluation, including the proposed target classifier, whose gradients can be back-propagated to the recognizer through a gradient reversal layer (GRL).}
    \label{fig:privacy_eval}
\end{figure}

\begin{figure*}[th]
\minipage{0.31\textwidth}
  \includegraphics[width=\linewidth]{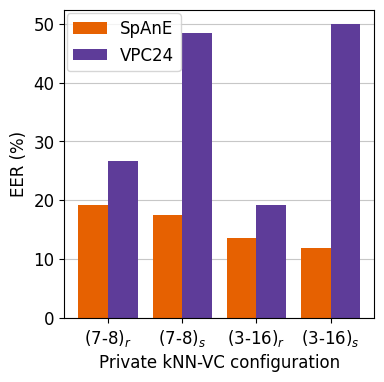}
  \caption{Different Private kNN-VC configurations evaluated with SpAnE and VPC24. The latter overestimates privacy for the same-gender TSA.}
  \label{fig:spane_vs_vpc}
\endminipage
\hfill
\minipage{0.32\textwidth}
  \includegraphics[width=\linewidth]{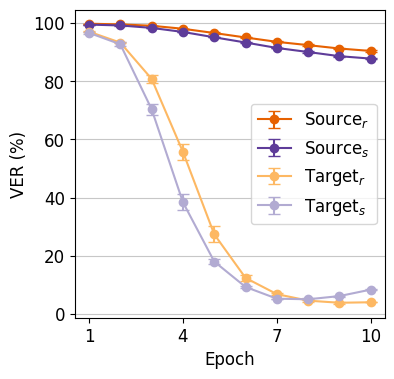}
  \caption{Classifier VERs per epoch for the baseline, for the two TSAs. The recognizer learns more about targets than about sources.}
  \label{fig:baseline_ver}
\endminipage
\hfill
\minipage{0.31\textwidth}
  \includegraphics[width=\linewidth]{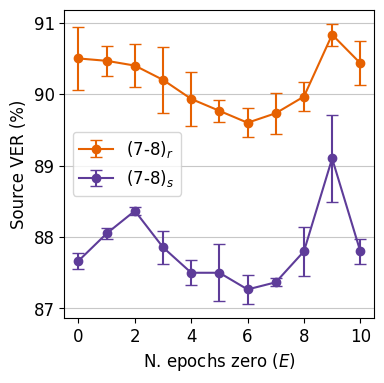}
  \caption{Final source VERs depending on for how many epochs $w_T$ is zero. The lowest VER is achieved for $E=6$, for both TSAs.}
  \label{fig:n_epochs_zero}
\endminipage
\hfill
\end{figure*}

\section{Privacy evaluation}

% privacy eval
We consider the evaluation protocol of the VoicePrivacy Challenge (VPC) 2024 \cite{tomashenko_voiceprivacy_nodate}, which proposes an attacker to identify anonymized speakers, called trial speakers.
The attacker has access to the anonymizer and to enrollment utterances from the trial speakers.
As depicted in Figure \ref{fig:privacy_eval}, the protocol comprises two steps: training and evaluation.
% Training
The attacker trains a speaker recognizer with data anonymized on \textit{utterance-level}, i.e. where a new target is randomly selected for every utterance, ensuring that the utterances of any speaker are anonymized with different targets.
A source classifier is appended to the recognizer, which predicts the source given the speaker embedding (the recognizer's output).
The recognizer is the ECAPA-TDNN implementation from SpeechBrain \cite{dawalatabad_ecapa-tdnn_2021}.

% Evaluation
The source classifier is removed for the evaluation, where the attacker's task is to identify the sources of the trial utterances, which are anonymized, by comparing them with the enrollment utterances, for which the attacker knows the sources.
The attack is strongest when both trial and enrollment utterances are anonymized on utterance-level, as was done for the training data \cite{champion_anonymizing_2023}.
The speaker embeddings of the enrollment utterances are averaged for each source.
They are then compared with the embeddings of the trial utterances with cosine similarity.
The metric for measuring the attack's performance based on the cosine similarities is the equal error rate (EER) \cite{maouche_comparative_2020}, where 0\% means all anonymized utterances were identified, and 50\% that the attacker was making binary decisions randomly, which implies perfect anonymization.

\subsection{Proposed architecture}

% our proposal
Our proposal is to add a second classifier during training, which, given the same speaker embedding that is passed to the source classifier, predicts the target.
It tells us how much target information the recognizer encodes.
Its gradients can be back-propagated to the speaker recognizer through a gradient reversal layer (GRL) \cite{ganin_unsupervised_2015}.
The GRL does nothing in the forward pass and negates the gradients in the backward pass.
This negation inverts the recognizer’s goal with respect to the target classifier: it will discard information that can be used by the target classifier to identify targets.
% GRL examples
The GRL has been used to make speech recognizers robust to accents \cite{sun_domain_2018}, and to make speech enhancement models robust to new kinds of noise unseen during training \cite{liao_noise_2019}.
Likewise, the proposed architecture intends to make the recognizer robust to targets.
We expect the recognizer to benefit from knowing the target, given its importance in characterizing anonymized speech.

\section{Experiment setup} \label{sec:setup}

% data
We use Librispeech \cite{panayotov_librispeech_2015} for our experiments: the \textit{train-clean-360} dataset is used for training and the \textit{test-clean} dataset is used for the evaluation.
10\% of the training data of each speaker is held out for validation at the end of each epoch.
The split between trial and enrollment utterances for the evaluation data is the one proposed in \cite{franzreb_optimizing_2025}; each of the 40 evaluation speakers has 20 utterances in each split.
For speakers with less than 40 utterances, the enrollment split is prioritized (only 9 utterances are missing from the expected 1,600).

% evaluation framework
We run our experiments with the SpAnE framework \cite{franzreb_comprehensive_2023}, which implements the VPC 2024 privacy evaluation.
Besides the evaluation dataset, the only difference with the VPC 2024 is the size of the recognizer: SpAnE uses the standard size provided by SpeechBrain \cite{dawalatabad_ecapa-tdnn_2021}, whereas the VPC 2024 uses a smaller one, with half the number of channels.
% Difference in gender
Also, the VPC 2024 evaluates the speakers of each gender separately, which makes the evaluation harder because there are no cross-gender comparisons \cite{franzreb_optimizing_2025}.
Our first experiment compares the two implementations.

% seeds
All experiments are repeated three times, each with a different initialization seed.
Error bars in plots and confidence scores in tables refer to the standard deviation across these seeds.
% Data anonymization
The training and evaluation data is only anonymized once for each anonymizer and TSA; it is the same across experiments and seeds.

\subsection{Private kNN-VC}

We run our experiments with private kNN-VC \cite{franzreb_private_2025}, as this was the anonymizer for which privacy was overestimated when the same-gender TSA was used.
It is an extension of kNN-VC \cite{baas_voice_2023}, meant to enhance its privacy by anonymizing prosodic aspects, which kNN-VC does not convert.
Phone durations and variations are anonymized to various degrees, depending on how it is configured.
We consider two configurations: $(7$-$8)$ and $(3$-$16)$.
The first parameter is the weight $w \in [0,1]$ of the predicted phone durations, which are interpolated with the actual durations, whose weight equals $1-w$.
It is multiplied by 10 in the configuration identifier for clarity.
The second parameter is the number of target candidates available for each phone.
Fewer candidates lower the information resolution available for the neighbor selection of kNN-VC, restricting phone variation.
From the two configurations, $(7$-$8)$ provides stronger anonymization of both phone durations and variations.

% Target selections
100 speakers of the LibriTTS \textit{train-other-500} dataset \cite{zen_libritts_2019} are used as targets, evenly distributed across genders.
We consider the two aforementioned TSAs (random and same-gender, stated as subscripts $r$ and $s$).

\section{Experiments}

We run three experiments.
The first one shows that using a larger recognizer solves the privacy estimation issues posed by the same-gender TSA.
The second experiment shows that the recognizer encodes more information about targets than sources, and the third experiment shows that adversarial learning removes target information from the recognizer, improving the evaluation's outcome for same-gender TSA.
The last experiment is performed with two further anonymizers, showing that the proposed evaluation generalizes well.

\subsection{Using a larger recognizer} \label{sec:exp1}

Our first experiment compares the privacy evaluations of the SpAnE and VPC 2024 frameworks.
As explained in Section \ref{sec:setup}, they differ in the size of the recognizer, the composition of the evaluation dataset and how gender is considered.
% Results
Figure \ref{fig:spane_vs_vpc} shows their EERs for the two private kNN-VC configurations and the two TSAs.
The VPC 2024 framework greatly overestimates privacy when the same-gender TSA is used: its EERs are close to 50\%, whereas they are much lower for the random TSA.
SpAnE's EERs are more consistent, suggesting better robustness to the TSA.
Its EERs are slightly lower for the same-gender TSA than for the random TSA, as the recognizer benefits from knowing the source speaker's gender.
For the random TSA, SpAnE's EERs are 7.5\% and 5.5\% lower for $(7$-$8)_r$ and $(3$-$16)_r$.

% Ablation study
To test whether the improvement comes from the larger recognizer or the evaluation dataset, we have evaluated the anonymizers with the same-gender TSAs, the VPC 2024 dataset and the larger recognizer.
The EERs are 22.4\% for $(7$-$8)_s$ and 17.0\% for $(3$-$16)_s$, meaning that the overestimation comes from using a smaller recognizer.

\subsection{Source and target information in the baseline}

We now measure the amount of information about sources and targets present in the speaker embeddings during training for $(7$-$8)_r$ and $(7$-$8)_s$.
The target classifier receives the speaker embeddings, exactly like the source classifier, but its gradients are not back-propagated to the recognizer.
% VERs
Figure \ref{fig:baseline_ver} depicts the validation error rates (VERs) of both classifiers per epoch.
The subscripts in the legend refer to the TSA.
The final avg. VERs for the source classifier are 90.4\% for the random TSA and 87.8\% for the same-gender TSA, whereas for the target classifier they are much lower: 3.9\% for the random TSA and 8.4\% for the same-gender TSA.
These results show that the recognizer is learning information that is more relevant for identifying targets than sources, although it is not informed about the targets in any way.

\subsection{Informing the recognizer about targets}

To improve the recognizer's performance, we now inform it about which target was used to anonymize each utterance by back-propagating the target classifier's gradients adversarially.
The influence of the target classifier on the recognizer is defined by its weight $w_T \in [0,1]$.
The study that proposed the GRL \cite{ganin_unsupervised_2015} gradually increases the $w_T$ after some epochs where it doesn't back-propagate to the backbone at all ($w_T = 0$), to suppress noisy signals at the beginning of training.
We implement this by increasing $w_T$ linearly from epoch $E$ until the last epoch (the 10th), where both classifiers have a weight of 1.
The source classifier's weight $w_S$ is always 1.

% n_epochs_zero
To decide on the value of $E$, i.e. for how long $w_T$ should be 0, we run $(7$-$8)$ for both TSAs and all possible values of $E$.
The resulting source VERs are depicted in Figure \ref{fig:n_epochs_zero}.
% Results
$E=10$ is the baseline, where the target classifier's gradients are never back-propagated to the recognizer.
For both TSAs, and including the baselines, the best VER is achieved for $E=6$.

% EERs
We consider two other anonymizers besides private kNN-VC for this final experiment: the original kNN-VC \cite{baas_voice_2023} and ASR-BN \cite{champion_anonymizing_2023}, the strongest baseline of the VPC 2024 challenge.
For kNN-VC, we use the same targets as for private kNN-VC.
ASR-BN provides 247 targets from the LibriTTS \textit{train-clean-100} dataset \cite{zen_libritts_2019}.
Table \ref{tab:eer_results} compares the results achieved by the baseline SpAnE and our proposal, where a target classifier is added with the aforementioned weight scheduling.
% Metrics
We consider three metrics: the EER and the VERs of the source and target classifiers.
A lower EER and source VER is better; it means that more anonymized speakers are identified.
On the other hand, a large target VER is better: it means that the recognizer is discarding target information.

\begin{table}[ht]
\centering
\caption{Comparison of our evaluation with SpAnE. All values depict means and standard deviations, as percentages.}
\label{tab:eer_results}
\begin{tabular}{llccc}
\toprule
\textbf{Anon.} & \textbf{Eval.} & \textbf{EER $\downarrow$} & \textbf{VER$_S$ $\downarrow$} & \textbf{VER$_T$ $\uparrow$} \\
\midrule

\multirow{2}{*}{\texttt{$(7$-$8)_r$}} & SpAnE & $19.1 \pm 0.5$ & $90.4 \pm 0.2$ & $4.0 \pm 0.0$ \\
& Ours & $19.4 \pm 1.2$ & $89.6 \pm 0.2$ & $99.4 \pm 0.0$ \\
\cmidrule(lr){2-5}
\multirow{2}{*}{\texttt{$(7$-$8)_s$}} & SpAnE & $17.4 \pm 0.3$ & $87.8 \pm 0.1$ & $8.4 \pm 0.1$ \\
& Ours & $15.9 \pm 0.9$ & $87.3 \pm 0.2$ & $99.5 \pm 0.1$ \\
\midrule
\multirow{2}{*}{\texttt{$(3$-$16)_r$}} & SpAnE & $13.6 \pm 1.0$ & $80.9 \pm 0.2$ & $10.4 \pm 0.5$ \\
& Ours & $12.4 \pm 0.2$ & $80.0 \pm 0.2$ & $99.3 \pm 0.0$ \\
\cmidrule(lr){2-5}
\multirow{2}{*}{\texttt{$(3$-$16)_s$}} & SpAnE & $11.8 \pm 0.3$ & $77.6 \pm 0.2$ & $17.6 \pm 0.7$ \\
& Ours & $10.2 \pm 0.4$ & $76.8 \pm 0.3$ & $99.3 \pm 0.0$ \\
\midrule
\multirow{2}{*}{\texttt{ASR-BN$_r$}} & SpAnE & $18.4 \pm 0.2$ & $92.1 \pm 0.1$ & $60.2 \pm 0.5$ \\
& Ours & $18.9 \pm 0.2$ & $91.9 \pm 0.1$ & $99.6 \pm 0.0$ \\
\cmidrule(lr){2-5}
\multirow{2}{*}{\texttt{ASR-BN$_s$}} & SpAnE & $17.4 \pm 0.6$ & $86.0 \pm 0.1$ & $37.3 \pm 0.4$ \\
& Ours & $13.9 \pm 0.4$ & $85.4 \pm 0.1$ & $99.5 \pm 0.0$ \\
\midrule
\multirow{2}{*}{\texttt{kNN-VC$_r$}} & SpAnE & $6.3 \pm 0.7$ & $40.9 \pm 0.3$ & $63.4 \pm 0.2$ \\
& Ours & $6.5 \pm 0.2$ & $41.4 \pm 0.2$ & $99.1 \pm 0.0$ \\
\cmidrule(lr){2-5}
\multirow{2}{*}{\texttt{kNN-VC$_s$}} & SpAnE & $5.2 \pm 0.1$ & $38.9 \pm 0.4$ & $63.6 \pm 1.1$ \\
& Ours & $5.0 \pm 0.2$ & $39.5 \pm 0.2$ & $98.8 \pm 0.0$ \\
\bottomrule
\end{tabular}
\end{table}

\iffalse
# Table insights:

- kNN-VC does not benefit from our evaluation, because it is not confused by the TSA at all.
    - It is the only anonymizer whose source VERs are lower than its target VERs, suggesting that it is not confused by the TSA.
    - kNN-VC's EERs are significantly lower than the rest.
    - The target VERs are the largest of all anonymizers.

Excluding kNN-VC from now on:

- The EER standard deviations for private kNN-VC are large.
    - TODO
- Random TSA does not benefit from our evaluation.
    - Agreeing with previous work that random TSA leads to a well-defined training dataset, where the recognizer is not confused by the TSA.
- Source VERs improve for all anonymizers and TSAs.
    - TODO
- Same-gender TSA greatly benefits from our evaluation.
    - the relative improvement in EER ranges from 9\% for $(7$-$8)_s$ to 20\% for ASR-BN$_s$.
    - ASR-BN is the only anonymizer whose random target VER is higher than its same-gender target VER for the baseline, suggesting that the recognizer is confused the most by the same-gender TSA, overly focusing on the targets.
- For all anonymizers and TSAs, the target VER is over 99\% for our evaluation.
    - Target information is successfully ignored by the recognizer.
\fi

% knn-vc
kNN-VC is by far the weakest among the four anonymizers ($EER < 6.5\%$), and the only one that does not benefit from our evaluation for the same-gender TSA.
It is also the only anonymizer whose source VER is lower than its target VER for the baseline.
Furthermore, its target VERs are the highest overall (over 63\%).
These results suggest that kNN-VC is not confused by the TSA, explaining why it does not benefit from our evaluation.

% random
For the random TSA, the evaluation does not benefit from the target classifier, agreeing with previous work that random TSA leads to a well-defined training dataset \cite{champion_anonymizing_2023, panariello_risks_2025}.
Training does perform slightly better for our evaluation, as it improves the source VERs for all anonymizers and TSAs.
% same-gender
For the same-gender TSA and the stronger anonymizers, our evaluation performs better.
The relative improvement in EER ranges from 9\% for $(7$-$8)_s$ to 20\% for ASR-BN$_s$.
% Target VERs
The largest difference between the two evaluations is naturally the target VER.
For our evaluation, it's always over 99\%: the recognizer successfully ignores target information.

% Private kNN-VC
Looking at the baseline evaluation, the private kNN-VC anonymizers encode the most target information, with target VERs ranging from 4.0\% for $(7$-$8)_r$ to 17.6\% for $(3$-$16)_s$.
For both configurations, the target VERs are higher when the same-gender TSA is used, and the EERs and source VERs are lower.
As discussed in Section \ref{sec:exp1}, the recognizer benefits from knowing the source gender, leading to a stronger attack on their privacy.
Adding the target classifier improves all VERs, and all EERs except for $(7$-$8)_r$, although its target VER for the baseline is the lowest.
The large standard deviations of the EERs (1.2\% for our evaluation) suggest that running more experiments would help clarify the behavior of this anonymizer's evaluation.

% ASR-BN's different baseline behavior
ASR-BN's baseline target VERs are considerably larger than those of private kNN-VC, whereas the baseline source VERs are comparable to those of $(7$-$8)$.
More importantly, its target VER for the same-gender TSA (37.3\%) is lower than the target VER for the random TSA (60.2\%).
For private kNN-VC, it was the other way around: the same-gender constraint improved the evaluation's performance and increased the target VER.
For ASR-BN, the same-gender constraint also improves performance, with lower source VERs and EERs, but the target VER decreases, suggesting that better evaluation performance is not always linked to a low target VER.
% ASR-BN benefits from our evaluation for same-gender TSA
Still, removing target information helps the recognizer to focus on identifying sources, reducing the mean EER from 17.4\% to 13.9\% for ASR-BN$_s$.

\section{Conclusion}

% Contribution
Our contribution aims to adapt the privacy evaluation, which currently uses the same architecture as the one designed for identifying original speakers, to deal with anonymized speech, which is radically different.
Our proposal to add a target classifier is simple to implement and does not increase the evaluation runtime significantly.
It improves our knowledge of the recognizer, showing whether it is being confused by the TSA.
If it is, we can remove target information from the recognizer with the target classifier.

% Results
Our study shows that using a larger recognizer improves the robustness of the privacy evaluation to the same-gender TSA, but that the evaluation's recognizer encodes a lot of target information.
Back-propagating the target classifier's gradients adversarially removes target information from the speaker embeddings, improving the recognizer's ability to characterize sources when the same-gender TSA is used.
The proposed evaluation is effective for two anonymizers, even though their representation of source and target information differ.
% Outlook
Future work should test the effectiveness of this approach for other anonymizers and TSAs.

\vfill\pagebreak

% References should be produced using the bibtex program from suitable
% BiBTeX files (here: strings, refs, manuals). The IEEEbib.bst bibliography
% style file from IEEE produces unsorted bibliography list.
% -------------------------------------------------------------------------
\bibliographystyle{IEEEbib}
\bibliography{strings,refs}

\end{document}